\documentstyle[preprint,aps,tighten,epsf,floats]{revtex}
\newcommand{\ben}{\begin{displaymath}}
\newcommand{\een}{\end{displaymath}}
\newcommand{\be}{\begin{equation}}
\newcommand{\ee}{\end{equation}}
\newcommand{\bea}{\begin{eqnarray}}
\newcommand{\eea}{\end{eqnarray}}

\newcommand{\fign}[1]{\label{#1}}

\begin{document}
\draft
\title{Issues of Regularization and Renormalization in
        Effective Field Theories of NN Scattering}
\author{ J. Gegelia}
\address{Department of Physics, The Flinders University of South Australia,
Bedford Park, SA 5042, Australia}
\date{\today}
\maketitle
              
\begin{abstract}
The equivalence of subtractively renormalized and cut-off effective field
theories is
demonstrated for the example of very low energy effective field theory for the
nucleon-nucleon interaction.

\end{abstract}

\pacs{
03.65.Nk,  
11.10.Gh,  
12.39.Fe,  
13.75.Cs.} 

\section{Introduction}

It is widely believed that $QCD$ is the correct theory of strong interactions.
On the other hand nuclear forces are studied within different potential
models. It is not clear whether or not these phenomenological approaches can be
justified from fundamental theory. Effective field theory is thought as a bridge
between $QCD$ and potential models. Chiral perturbation theory serves as a
low-energy effective field theory inspired by QCD.   

There has been much recent interest in the EFT approach to the nucleon-nucleon
scattering problem (see 
\cite{ordonez,cohen1,phillips1,scaldeferri,phillipswsh,unp,beane,kaplan,kaplan2,kaplan3,gegelia,gegelia1,park}
and citations in these papers).
The chiral perturbation theory approach for processes involving
an arbitrary number of nucleons was formulated in
\cite{weinberg1,weinberg2}. Unlike  purely pionic processes
\cite{wein1979}, for the
$n$-nucleon 
problem  power counting
is used for the ``effective potentials''  instead of full amplitudes.
The effective potential is defined as a sum of time-ordered perturbation theory
diagrams for the $T$-matrix excluding those with purely nucleonic intermediate
states.  

 To find the full $S$-matrix one should solve a Lippmann-Schwinger equation
(or Schr\" oedinger equation) with this effective potential in place of the
interaction Hamiltonian, and with {\it only} $n$-nucleon intermediate states 
 \cite{weinberg1}.

  The Lagrangian of effective field theory  is highly
non-renormalizable in the traditional sense but it contains all possible terms
which are not suppressed by the symmetries of the theory and the ultraviolet
divergences  are
absorbed into the parameters of the Lagrangian.
Renormalization points are chosen of the order of external
momenta $p$.
After renormalization, the effective cut-off is of order $p$  \cite{weinberg2}. 
Power counting apparently depends on the normalisation condition (One could
choose a normalisation condition for which the power counting breaks down for
extremally low energies or there is no power counting at all). If one wants the
power counting to be valid for high enough energies, one should use an
appropriately chosen normalisation condition. While the complete expressions of
physical quantities  
should not depend on this choice the renormalised Feynman diagrams certainly
do.

Subtractively renormalised effective field theory encounters quite severe
(technical) problems:
if one takes the potential up to some order and tries to iterate  via
the Lippmann-Schwinger equation one will encounter divergences. One could try
to include counter-terms in the potential, but due to the
non-renormalizability of the
theory the inclusion of an infinite number of terms with more and more
derivatives would be needed. One could even think that Weinberg's power counting
breaks down
because higher order counter-terms are strongly involved. But it should be
remembered
that 
power counting (for both amplitudes and the potentials) is valid after
renormalization when the contributions of counter terms are already taken into
account \cite{weinberg1,weinberg2,gegelia}. As was explained in details in
\cite{gegelia1} and \cite{park} this involvement of higher order counter-terms
into low order calculations do not invalidate Weinberg's power counting 
arguments.  
 So, one has either to exactly solve (formally) the equation and after subtract
divergences explicitly, or otherwise one should draw all relevant diagrams,
subtract them and then sum these renormalised diagrams up. In recent papers
\cite{kaplan,kaplan2,kaplan3} Kaplan et. al suggested a systematic method of
summation of an infinite number of diagrams using dimensional regularization and
the Power Divergent Subtraction scheme. But, as was mentioned in the above cited
papers, in the theory with explicitly included pions for the external momenta
exceeding 100 MeV it is difficult to justify
suggested approximations (in particular the perturbative inclusion of pions).
So for higher energies the problem of summation of renormalized diagrams remains
open.   
    Fortunately these problems can be overcome using cut-off theory. One can
calculate up to any desired order, but there is a very crucial question: what
is the relation between subtractively renormalised and  cut-off theories? 
This question is addressed in a number of papers
\cite{phillipswsh,unp,beane,kolck22,kolck07,adhikari,lepage,lepage1}, but as yet
the 
complete answer has not been determined. Moreover some authors question the
validity and systematic character of
cut-off theory calculations (see for example
\cite{phillipswsh,epelbaoum,epelbaoum1}). This work is an attempt to investigate
some details about the equivalence of subtractively renormalized and cut-off
theories.

If one applies Weinberg's power counting directly to the
amplitude of $NN$ scattering one gets 
a series:
\begin{equation}
T=T_0+T_1+T_2+T_3+...
\label{nuexpampl}
\end{equation}
where $T_i$ is suppressed by $i$-th order of a small parameter. Each term in
this infinite series is a sum of an 
infinite number of diagrams itself. These diagrams are of the same order. If
translated into the language of the potential, $T_0$ contains all diagrams which
are obtained when leading order potential is iterated via the Lippmann-Schwinger
equation. $T_1$ contains diagrams with one insertion of the first order
potential and arbitrary number of the leading order potential. $T_2$ contains
all diagrams with one insertion of second order diagram or two insertions of
first order potential etc. (Note that for a theory with just nucleons, which is
considered in this paper, all $T_j$ vanish for odd $j$.) The
expansion parameter in (\ref{nuexpampl}) is $\sim Q/\Lambda$ where $Q$ is of
the order of external momenta
and $\Lambda $ is expected to be of the order of the mass of lightest particle
which was integrated out. If
$Q<<\Lambda$ then the first few terms of 
(\ref{nuexpampl}) should approximate the whole amplitude well. If applied to    the complete theory the equivalence of
subtractively renormalised and cut-off theories would require that they produce
identical series for scattering amplitudes provided that
the same  normalisation condition is used.  

In actual calculations one takes a potential up to some order and solves
Lippman-Schwinger equation. Scattering amplitude determined from
Lippman-Schwinger equation contains all contributions up to given order (the
order of the potential) and
also some parts of higher order contributions i.e. some diagrams contributing to
$T_j$ with large $j$ are included and others are not. As far as these
contributions are 
small it is not inconsistent to include part of higher order contributions
while other parts are not included; the error is of the order of neglected
terms.  Beyond the range of validity of power
counting high order contributions become large and make the complete expression
unreliable.      
As for the equivalence of subtractively renormalised and cut-off theories
one would 
expect that they will give identical results up to the order of considered
approximation. The  
difference between two results should be small, of the order of neglected
terms. Of course one would not expect that this difference is small beyond the
range of validity of power counting. 

Eq. (\ref{nuexpampl}) generates an expansion in the same small parameter for the
inverse amplitude 
$$
{1\over T}={1\over T_0}-{T_1\over T_0^2}-{T_0T_2-T_1^2\over T_0^3}+\cdots
$$  
If two expressions for the amplitude generated by subtractively renormalised and
cut-off theories are equal up to some order and the difference between 
them is small then the same is true for inverse
amplitudes and vice-versa.

    Below the simple example of contact interaction of nucleons in
$^1S_0$ wave is considered. The amplitude is renormalized by subtracting
divergent integrals at
some normalisation point and  its relation to the amplitude obtained from
cut-off theory is studied. The numerical values of 
phase shifts obtained from subtractively renormalised and  
cut-off theories (without removing cut-off) are compared.


\medskip
\medskip
\medskip

\section{$p^2$ order calculations}

For the very low energy nucleon-nucleon scattering processes the
pions can be integrated  out 
and the effective non-relativistic Lagrangian takes the following form
\cite{kaplan}: 
$$
{\cal L}=N^{\dagger}i\partial_tN+N^{\dagger}{\nabla^2 \over 2M}N-
{1\over 2}C_S\left( N^{\dagger}N\right)^2-{1\over 2}C_T
\left( N^{\dagger}\mbox{\boldmath $\sigma$}N\right)^2
$$
\begin{equation}
-{1\over 2}C_2\left( N^{\dagger}\nabla^2 N\right)
\left( N^{\dagger}N\right)+h.c.+...
\label{e1}
\end{equation}
where the nucleonic field $N$ is a two-spinor in spin space and a two-spinor in
isotopic spin space and $\mbox{\boldmath $\sigma$} $ are the Pauli matrices
acting on spin indices. $M$ is
the mass of nucleon and the ellipses refer to additional 4-nucleon operators
involving two or more derivatives, as well as relativistic corrections to the
propagator. $C_T$ and
$C_S$ are couplings introduced by Weinberg \cite{weinberg1,weinberg2}, they are
of dimension $(mass)^{-2}$ and $C_2$ is of the order $(mass)^{-4}$. 

The leading order contribution to the 2-nucleon potential is

\begin{equation}
V_0\left( {\bf p},{\bf p'}\right)=C_S+
C_T\left( \mbox{\boldmath $\sigma$}_1,\mbox{\boldmath $\sigma$}_2\right).
\end{equation}
In the ${ }^1S_0$ wave it gives:
\begin{equation}
V_0\left( {p},{p'}\right)=C
\end{equation}
where $C=C_S-3C_T$.
The next to leading order contribution to the 2-nucleon
potential in the ${ }^1S_0$ channel takes the form: 

\begin{equation}
V_2\left( {p},{p'}\right)=C_2\left({p}^2+{p'}^2\right)
\end{equation}

The formal iteration of the potential $V_0+V_2$ using the Lippmann-Schwinger
equation gives for on-shell ($E=p^2/M$) $s$-wave 
$T$-matrix \cite{unp}: 
 
\begin{equation}
{1\over T(p)}={\left( C_2I_3-1\right)^2\over C+C_2^2I_5+
p^2C_2\left( 2-C_2I_3\right)}-I(p)
\label{2}
\end{equation}
\begin{equation}
I_n=-M\int {d^3k\over (2\pi )^3}k^{n-3}; \ \ 
I(p)=M\int {d^3k\over (2\pi )^3}{1\over p^2-k^2+i\eta}=I_1-{iMp\over 4\pi},
\label{3}
\end{equation}
where $p$ is the on-shell momentum and $I_1$, $I_3$ and $I_5$ are divergent
integrals.

\subsection{Renormalization by subtracting divergences}
To renormalize (\ref{2}) it is necessary to include contributions of an infinite
number of counter-terms with higher and higher (up to infinity) derivatives 
\cite{gegelia}. While it is impossible to write down all these contributions
explicitly it is quite straightforward to renormalize (\ref{2}) by just
subtracting divergent integrals. Before implementing this scheme it would be
useful to write down some of the leading and $p^2$ order counter-terms. 

One can write down the chiral expansion for $T$-matrix \cite{kaplan} (it is
equivalent to an expansion of $T$ obtained from (\ref{2}) in powers of $C_2$):
\begin{equation}
T={C\over 1-CI(p)}+{2p^2C_2+2CC_2I_3\over \left( 1-CI(p)\right)^2}+...
\label{chtmatrix}
\end{equation}   
and 
\begin{equation}
{1\over T(p)}=-I(p)+{1-2C_2I_3\over C}-{2C_2p^2\over C^2}+...
\label{invchtm}
\end{equation}   

The final goal is to absorb divergences in (\ref{chtmatrix}) into $C$ and $C_2$
which are to be given by 
\begin{equation}
C=C^{(1)}\left( C_R\right)+C_2^RC^{(2)}\left( C_R\right)+...
 \ \ \ \ C_2=C_2^RC_2^{(1)}\left( C_R\right)+...
\label{cterms}
\end{equation}  
where $C_R$ and $C_2^R$ are renormalized coupling constants and 
$C^{(1)}\left( C_R\right)$, $C^{(2)}\left( C_R\right)$, 
$C_2^{(1)}\left( C_R\right)$... are functions of $C_R$.

To determine $C$ and $C_2$ in terms of $C_R$ and $C_2^R$ 
it is simpler to work with (\ref{invchtm}) and require:
\begin{equation}
{1-2C_2I_3\over C}-{2C_2p^2\over C^2}-
I(p)={1-2C_2^R\left( I_3-\Delta_3\right)\over C_R}-{2C_2^Rp^2\over C^2_R}-
(I(p)-\Delta )
\label{npct}
\end{equation}
Where $\Delta$ and
$\Delta_3$ are divergent parts of $I(p)$ and $I_3$ integrals (with arbitrary
finite contributions).
Equating coefficients of different powers of $p$ one gets from (\ref{npct}):
\begin{equation}
{1-2C_2I_3\over C}={1-2C_2^R\left( I_3-\Delta_3\right)\over C_R}+\Delta ; \ \ \
\  {C_2\over C^2}={C_2^R\over C^2_R}
\label{npct1}
\end{equation}
and from (\ref{npct1}):
\begin{equation}
C={-C_R^2\Delta -C_R\left\{ 1-2C_2^R\left
( I_3-\Delta_3\right)\right\}\pm C_R\left( 8C_2^RI_3+\left[ 1-2C_2^R\left
( I_3-\Delta_3\right) +C_R\Delta\right]^2\right)^{1\over 2}\over 4C_2^RI_3}
\label{npct2}
\end{equation}
In ordinary perturbation theory (expansion in terms of coupling constants) one
has $C=C_R+...$. The non-perturbative expression (\ref{npct2}) respects this
perturbative expansion if the $''+''$ sign is taken. Expanding the 
chosen solution in powers of $C^R_2$ and keeping only terms of first order one
obtains:
\begin{equation}
C={C_R\over 1+C_R\Delta}+{2C_RC_2^R\left( I_3-\Delta_3\right)\over 
\left( 1+C_R\Delta\right)^2}-{2C_RC_2^RI_3\over 
\left( 1+C_R\Delta\right)^3}
\label{npct3}
\end{equation}
and
\begin{equation}
C_2={C_2^R\over \left( 1+C_R\Delta\right)^2}
\label{npct4}
\end{equation}
Substituting (\ref{npct3}) and (\ref{npct4})  into (\ref{chtmatrix}) one gets
a finite renormalized expression:
\begin{equation}
T={C_R\over 1-C_R\left[ I(p)-\Delta\right]}+{2p^2C_2^R+
2C_RC_2^R\left( I_3-\Delta_3\right)\over \left( 1-C_R\left[ I(p)-
\Delta\right]\right)^2}+...
\label{renchtmatrix}
\end{equation}
So, to get rid off divergences from (\ref{chtmatrix}) one has just to express
bare 
couplings in terms of renormalised ones using (\ref{npct3}) and (\ref{npct4})
and substitute into (\ref{chtmatrix}). One gets an expression with subtracted
integrals and renormalised coupling constants. 
 
Switching back to (\ref{2}) one can apply the subtraction scheme analogous
to the one originally used by
Weinberg \cite{weinberg2} and subtract divergent integrals at $p^2=-\mu^2$.
Integrals are divided into two parts:
\begin{equation}
I_n=I_n\left( p^2=-\mu^2\right)+\left[ I_n-I_n\left( p^2=-\mu^2\right)\right]=
I_n^d+I_n^R, \
n=3,5 
\end{equation}
\begin{equation}
I(p)=I\left( p^2=-\mu^2\right)+\left[ I(p)-I\left( p^2=-\mu^2\right)\right]=
I^d+I^R(p)
\end{equation}
where $I_n^d$ and $I^d$ are divergent parts and are to be cancelled by
contributions of counter-terms.  To absorb
all contributions of 
$I_n^d$ and $I^d$ in (\ref{2}) one needs to include contributions of an infinite
number of counter-terms with higher and higher order (up to infinity)
derivatives \cite{gegelia}. While it is impossible to write down
these counter terms explicitly, one can take their contributions into account by
just neglecting $I_n^d$ and $I^d$ terms and replacing $C$ and $C_2$ by
renormalized couplings. 
Finally the amplitude is left with
finite parts of integrals $I_n^R$ and $I^R(p)$ (note that $I_n^R=0$):
\begin{equation}
{1\over T}={1\over C_R+2C^R_2p^2}-I^R(p)
\label{rentmp2}
\end{equation}
where
\begin{equation}
I^R(p)=-{M\over
4\pi}\mu-{M\over 4\pi}ip
\label{renint1}
\end{equation}

Note that the subtraction
scheme used here is just one among an infinite number of possibilities. Any
scheme which puts effective cut-off of the order of external momenta should be
equally good.   

Eq. (\ref{rentmp2}) is not obtained from (\ref{2}) by just expressing $C$ and
$C_2$ in terms of renormalised coupling constants. Contributions of infinite
number of counter-terms have been taken into account. While the imaginary part
of (\ref{2}) can not be altered by adding contributions of counter-terms, the
real part depends on finite parts of those counter-terms. Hence starting from
the same (inverse) amplitude in terms of bare parameters one can get quite
different expressions for renormalised (inverse)amplitude. All these amplitudes
are equivalent up to the order one is working with provided that chosen
normalisation conditions respect power counting.    

Note that although the expression (\ref{rentmp2}) was obtained in
\cite{kaplan2} using Power Divergent Subtractions, that scheme is completely
different from the one applied in this work. The difference is clearly seen
when pions
are included explicitly. 

Matching (\ref{rentmp2}) to the effective range expansion 
\begin{equation}
{1\over T}=-{M\over 4\pi}\left( -{1\over a}+{1\over 2}r_ep^2+
0\left( p^4\right)-ip\right)  
\label{effre}
\end{equation}
gives
\begin{eqnarray}
{1\over C_R}+{M\over 4\pi}\mu={M\over 4\pi a} 
\label{matchingc}
\\
{2C_2^R\over C_R^2}={Mr_e\over 8\pi} 
\label{matchingc2}
\end{eqnarray} 
and from (\ref{rentmp2}) 
\begin{equation}
{1\over T}=-{M\over 4\pi}\left( {-{1\over a}(1-a\mu )-{1\over 2}r_ea\mu p^2\over
1-a\mu +{1\over 2}r_ea p^2}-ip\right)
\label{rentmp2ere}
\end{equation}
The result given in (\ref{rentmp2ere}) does not depend on the regularization
scheme.

Below a cut-off version of the above effective theory is considered and it is
demonstrated
that these two approaches are equivalent up to (including) $p^2$ order. 
\subsection{Cut-off theory}

Effective potential with sharp cut-off has the following form:
\begin{equation}
V^{(2)}(p',p)=\left\{ \bar{C}+\bar C_2\left( p^2+
p'^2\right)\right\}\theta (l-p)\theta (l-p')
\label{V_R}
\end{equation}
Here $l$ is the cut-off parameter and $\bar C$ and
$\bar C_2$ depend
on this parameter. $l$ should be of the order of the mass of
lightest particle which was integrated out \cite{lepage}. 
It is not difficult to write down the solution of the Lippmann-Schwinger
equation explicitly (see \cite{beane}):

\begin{equation}
\frac{1}{T(p)}=\frac{(\bar C_2 I_3^l -1)^2}{\bar C+\bar C_2^2 I_5^l + {p^2}
\bar C_2 (2 -
\bar C_2 I_3^l)} - I_l(p),
\label{T2}
\end{equation}
where
\begin{equation}
I_n^l \equiv -M \int \frac{d^3q}{(2 \pi)^3} q^{n-3}\theta (l-q)=
-{M\over 2\pi^2}{l^n\over n}
\label{In}
\end{equation}

\begin{equation}
I_l(p)=M \int \frac{d^3q}{(2 \pi)^3} 
\frac{1}{p^2 - q^2+i\epsilon}
\theta( l - q)=-{M\over 2\pi^2}\left[ l+{p\over 2}ln{1-{p\over l}\over 
1+{p\over l}}\right]-{iMp\over 4\pi }
\label{In1}
\end{equation}
Matching (\ref{T2}) to the effective range expansion (\ref{effre}) leads to:
\begin{eqnarray}
{\left( 1-\bar C_2I_3^l\right)^2\over \bar C+\bar C_2^2I_5^l}
={M\over 4\pi a}-{Ml\over 2\pi^2}\equiv x 
\\ 
{\bar C_2\left( 2-\bar C_2I_3^l\right)\over \bar C+\bar C_2^2I_5^l}
={1\over x}\left( {Mr_e\over 8\pi}-
{M\over 2\pi^2l}\right)\equiv y
\end{eqnarray}    
and the solution of these equations for $\bar C$ and $\bar C_2$ gives:
\begin{eqnarray}
\bar C_2={1\over I_3^l}\left[ 1-\left( {x\over x+yI_3^l}\right)^{1\over 2}
\right] 
\label{c2cut} \\
\bar C={1\over x+yI_3^l}-\bar C_2^2I_5^l
\label{ccut}
\end{eqnarray}
$\bar C_2$ was obtained by solving quadratic equation. Analogously to
(\ref{npct2}) the sign in solution was fixed respecting the structure
of ordinary perturbation
theory (expansion in coupling constants). 

Substituting (\ref{c2cut}) and (\ref{ccut}) into (\ref{T2}) one gets:
\begin{equation}
{1\over T}=-{M\over 4\pi}\left( {-\pi +{2al}\over \pi a+
p^2{\pi a^2\left( \pi r_el-4\right)\over 2l\left( \pi-2al\right)}}+
{4\pi\over M}I_l(p)\right)
\label{T2efe}
\end{equation}
Note that higher-order corrections to the cut-off expression are suppressed by
powers of $p/l$ and hence are small for momenta well below the cut-off.

\begin{figure}[t]
\hspace*{5cm}  \epsfxsize=8cm\epsfbox{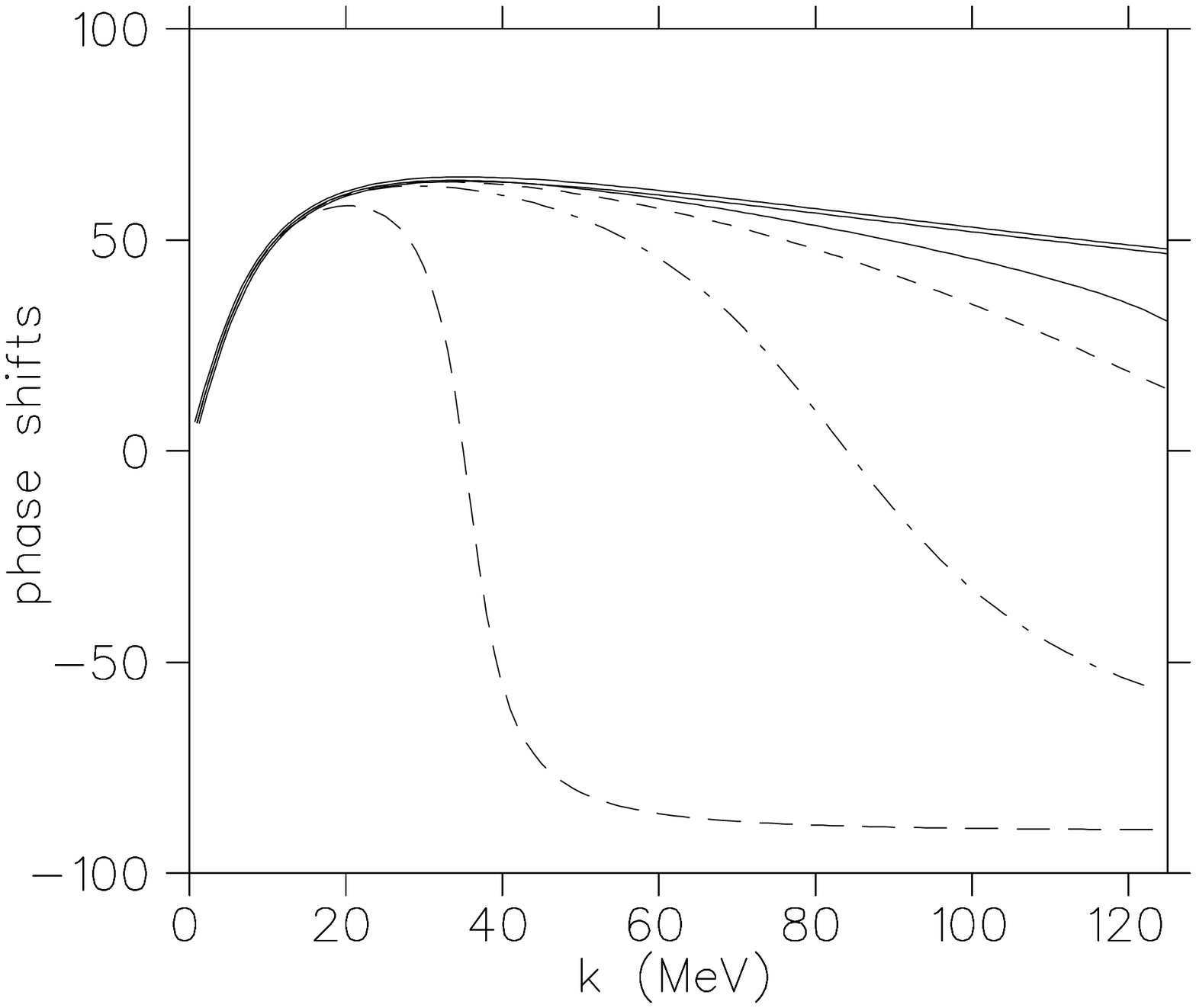}
\vspace{2mm}
\caption{\fign{phi3d}{\it Phase shifts calculated in $p^2$ order. Double line
corresponds to the effective
range expansion, solid line corresponds to the cut-off theory with $l=140 {\rm
MeV}$ and long-dashed,
dash-doted and short-dashed lines correspond to $\mu =0$, $\mu =40$ and
$\mu =130 \ {\rm MeV}$ respectively}}
\end{figure}

The solution of $a$ and $r_e$ from (\ref{matchingc}) and
(\ref{matchingc2}) (for some value of $\mu$) and substitution into
(\ref{c2cut}) and (\ref{ccut}) leads to
a lengthy but simple relation between $\bar C$, $\bar C_2$ and
$C_R$, $C_2^R$ the fulfilment of which guarantees that the cut-off and
subtractively renormalised inverse $T$-matrices   are equal up to
(including)
$p^2$ order.
This equality is manifested by (\ref{rentmp2ere})
and (\ref{T2efe}). Consequently in terms of $\nu$ expansion of the Feynman
amplitude given in \cite{kaplan} the two amplitudes are equal up to (including)
sub-leading
order. Higher order corrections to the cut-off expression are 
suppressed by powers of cut-off parameter $l$ which should be taken of the order
of lightest integrated particle, so they are of the order of terms which are
neglected by the approximation taken in renormalized theory.

\begin{figure}[t]
\hspace*{5cm}  \epsfxsize=8cm\epsfbox{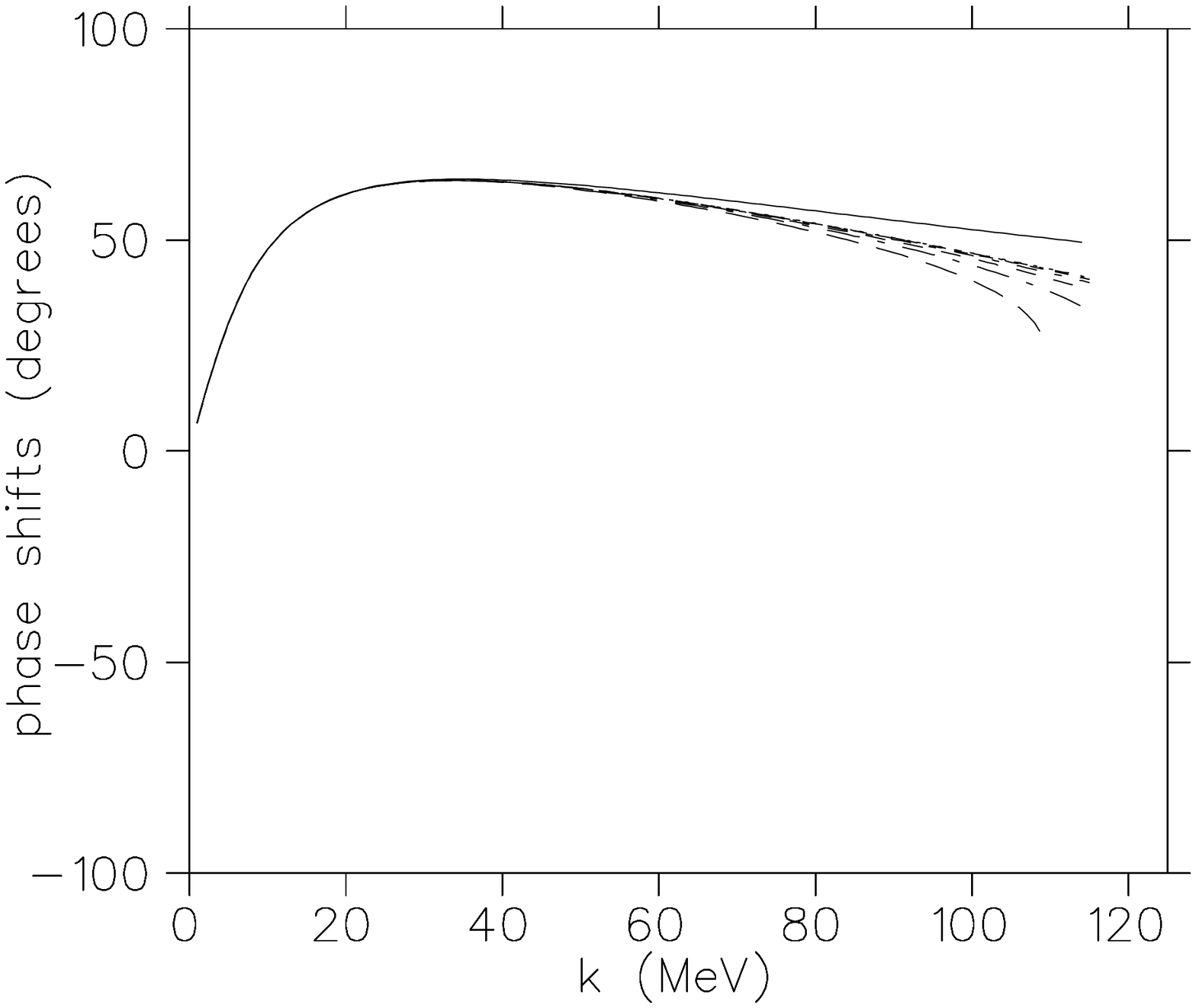}
\vspace{2mm}
\caption{\fign{phind}{\it Phase shifts calculated in $p^2$ order cut-off
theory. Solid line corresponds to the effective range expansion. The lowest
dashed line corresponds to cut-off parameter $l=110 \ {\rm MeV}$ and subsequent
lines correspond to the values
$l=120, \ 130, \ 140, \ 150, \ 160, \ 170 \ {\rm MeV}$.}}
\end{figure}

Substituting actual values for scattering length and effective range 
$a=-1/(8.4 \ {\rm MeV})$ and $r_e=0.0137 \ {\rm MeV}^{-1}$ into
(\ref{rentmp2ere}) and (\ref{T2efe}) one can
calculate the phase shifts. The
results for $l=130 \ {\rm MeV}$ and $\mu=0, 40, 130 \ {\rm MeV}
$
(Note that
$\mu =130 \ {\rm MeV}$ does not violate the power counting at least for the
present
problem) are plotted in FIG.1. As is seen from this graph if one takes $\mu\sim
100 \ {\rm MeV}$ the cut-off phase
shifts are in good  agreement with the ones of subtractively renormalised
theory for the momenta up to $\sim 80 \ {\rm MeV}$. If smaller value of $\mu$ is
taken then the range of the validity of power counting decreases. So does the
range of the momenta for which the results of two approaches are in good
agreement.   
As was mentioned above the
power counting depends on the chosen normalisation condition. One does not
expect good agreement between results of subtractively renormalised and cut-off
theories unless good normalisation condition is chosen. 
The $\mu =0$ graph shows the failure of $\overline {MS}$
renormalised theory encountered in \cite{kaplan}. One can also
calculate the numerical values of cut-off coupling constants $\bar C 
\approx -1/(78.2 \ {\rm MeV})^2$ and 
$\bar C_2 \approx 1/(155.5 \ {\rm MeV})^4$.

To study the dependence of phase shifts on cut-off parameter  the phase shifts
for different values of this parameter are
plotted in FIG.2. It is seen that for cut-offs $\sim m_\pi$ phase shifts do not
depend on cut-off up to momenta $\sim 50 \ {\rm MeV}$.

Note that figures quite analogous to FIG.1 and FIG.2 but in different context
and with different subsequent conclusions are given in
\cite{phillipswsh}.

\section{$p^4$ order calculations}

To estimate the corrections from the next orders let us consider
the $p^4$ order potential:
\begin{equation}
V\left( p,p'\right)= C+C_2\left( p^2+p'^2\right)+B\left( p^4+p'^4\right)+
B_1p^2p'^2
\label{potmp4}
\end{equation}

This potential can be written as separable one:
\begin{equation}
V\left( p,p'\right)=\sum_{i,j=0}^2p'^{2i}\lambda_{ij}p^{2j}
\label{seppotmp4}
\end{equation}
where
\begin{equation}
\{\lambda_{ij}\}_{i,j=0}^1=\left( \begin{array}{ccc} C & C_2 & B \\
						    C_2 & B_1 & 0 \\
						    B & 0 & 0	
				  \end{array} \right).
\label{lambda}
\end{equation}
The relativistic corrections are suppressed by the mass of
the nucleon
and 
hence they are not included.
A straightforward generalisation of calculations with $p^2$-order
potential given in \cite{unp} leads to the following expression:
\begin{equation}
{1\over T}={N\over D}-I(p)
\label{tmp4}
\end{equation}
where
$$
N=1-I_3\left( {2C_2}+{2p^2B}+{p^2B_1}\right)-I_5\left({2B}+{B_1}\right)
+I_5^2\left({2BB_1}+{B^2}\right)
$$
$$
+I_3^2\left( {p^4B^2}+{2C_2p^2B}+{C_2^2-CB_1}
\right)+I_3I_5\left( {2p^2BB_1}+{2p^2B^2}+{2C_2B}\right)-{2BB_1}I_3I_7
$$
\begin{equation}
-B^2B_1I_3^2I_9+p^2B^2B_1I_3^2I_7-B^2B_1I_5^3-B^2B_1p^2I_3I_5^2+2B^2B_1I_3I_5I_7
\label{t3num}
\end{equation}
and
$$
D=C+2C_2p^2+2p^4B+p^4B_1-I_5\left( 2p^4BB_1+p^4B^2+CB_1-C_2^2\right)
$$
$$
+I_3\left[ \left( CB_1-C_2^2\right)p^2-2C_2p^4B-p^6B^2\right]+
I_7\left( {2p^2BB_1}+{2p^2B^2}+{2C_2B}\right)+B^2I_9
$$
\begin{equation}
+p^2B^2B_1I_3I_9-p^4B^2B_1I_3I_7-p^2B^2B_1I_5I_7-B^2B_1I_5I_9+
p^4B^2B_1I_5^2+B^2B_1I_7^2
\label{t3denum}
\end{equation}

\subsection{Renormalization by subtracting divergences} 
Analogously to the $p^2$ order case one can renormalize (\ref{tmp4}) by
subtracting divergent integrals at $p^2=-\mu^2$ and get:
\begin{equation}
{1\over T}={1\over C_R+2C^R_2p^2+2p^4B^R+p^4B^R_1}-I^R(p)
\label{rentmp4}
\end{equation}
where $C_R$, $C^R_2$, $B^R$ and $B^R_1$ are renormalised coupling constants and 
\begin{equation}
I^R(p)=-{M\over
4\pi}\mu-{M\over 4\pi}ip
\label{renint}
\end{equation}
Comparing (\ref{rentmp4}) to the effective range expansion 
\begin{equation}
{1\over T}=-{M\over 4\pi}\left( -{1\over a}+{1\over 2}r_ep^2+
dp^4+0\left( p^6\right)-ip\right)  
\label{effrep4}
\end{equation}
gives
\begin{eqnarray}
{1\over C_R}+{M\over 4\pi}\mu={M\over 4\pi a} \label{a4r}
\\
{2C_2^R\over C_R^2}={Mr_e\over 8\pi} \label{r4r}
\\ 
{4\left( C_2^R\right)^2\over C_R^3}-{2B^R+B_1^R\over C_R^2}=-{Md\over 4\pi}
\label{matching}
\end{eqnarray}

\subsection{Cut-off theory}

Introducing a sharp cut-off  (factor of
$\theta (l-p)\theta (l-p')$) into the
potential (\ref{potmp4}) and solving Lippmann-Schwinger equation for the
$T$-matrix
one gets the expressions (\ref{tmp4}), (\ref{t3num}), (\ref{t3denum}) with $I_n$
replaced by $I_n^l$, $I(p)$ replaced by $I_l(p)$ and cut-off dependent
couplings $\bar C$, $\bar
C_2$, $\bar B$ and $\bar B_1$.

For the purposes of this work one can take
$\bar B=0$.  While simplifying calculations significantly this value is quite
satisfactory as far as adjusting remaining parameters one can satisfy the
equivalence criteria. Substituting this value one gets:

\begin{equation}
{1\over T}={\left( 1-\bar C_2I_3\right)^2-I_5{\bar B_1}
-I_3^2{\bar C\bar B_1}-I_3{\bar B_1}p^2\over \bar C-I_5\left( \bar C\bar B_1-
\bar C_2^2\right)+I_3\left( \bar C\bar B_1-\bar C_2^2\right)p^2+2\bar C_2p^2+
p^4\bar B_1}-I_l(p)
\label{b0t3}
\end{equation}
Comparing (\ref{b0t3}) to the effective range expansion (\ref{effrep4}), after
a lengthy but straightforward calculation one obtains:
\begin{eqnarray}
\bar B_1={y\over x+I_3z} \label{matchingcutpar1} \\
\bar C_2={1\over I_3}\left\{ 1-\bar B_1I_5\pm \left[ \left( 1-
\bar B_1I_5\right)^2-
{I_3z\over x+I_3z}+\bar B_1{I_3^2+zI_3I_5\over I_3x+
I_3^2z}\right]^{1\over 2}\right\}
\label{matchingcutpar2} \\
\bar C=-{2I_5\bar C_2\over I_3}+{I_3+zI_5\over I_3x+I_3^2z}
\label{matchingcutpar3}
\end{eqnarray} 
where 
\begin{eqnarray}
x=a_1+I_5{a_2^2+a_1a_3\over a_1^2+a_2I_3} \label{x} \\
y={a_2^2+a_1a_3\over a_1^2+a_2I_3} \label{y} \\
z={a_1a_2-a_3I_3\over a_1^2+a_2I_3} \label{z}
\end{eqnarray}
and
\begin{eqnarray}
a_1={M\over 4\pi a}-{Ml\over 2\pi^2} \label{a1} \\
a_2={Mr_e\over 8\pi}-{M\over 2\pi^2l} \label{a2} \\
a_3={Md\over 4\pi}-{M\over 6\pi^2l^3}
\label{a3}
\end{eqnarray}

Solving for $a$, $r_e$ and $d$ from (\ref{a4r})-(\ref{matching}) and
substituting
into (\ref{matchingcutpar1})-(\ref{a3}) one obtains lengthy algebraic relations
between
$\bar C$, $\bar C_2$, $\bar B_1$ and $C_R$, $C_2^R$, $B^R$ and $B_1^R$, the
fulfilment of which along the condition $\bar B=0$ guarantees the equality of
cut-off and subtractively renormalized inverse amplitudes  up
to (including) $p^4$ order, and consequently the $T$ matrices of two approaches
are
equal up to (including) $\nu =4$ order. The higher order corrections to the
cut-off
expression are again suppressed by powers of $p/l$ and hence are small for
momenta well below the cut-off.

\begin{figure}[t]
\hspace*{5cm}  \epsfxsize=8cm\epsfbox{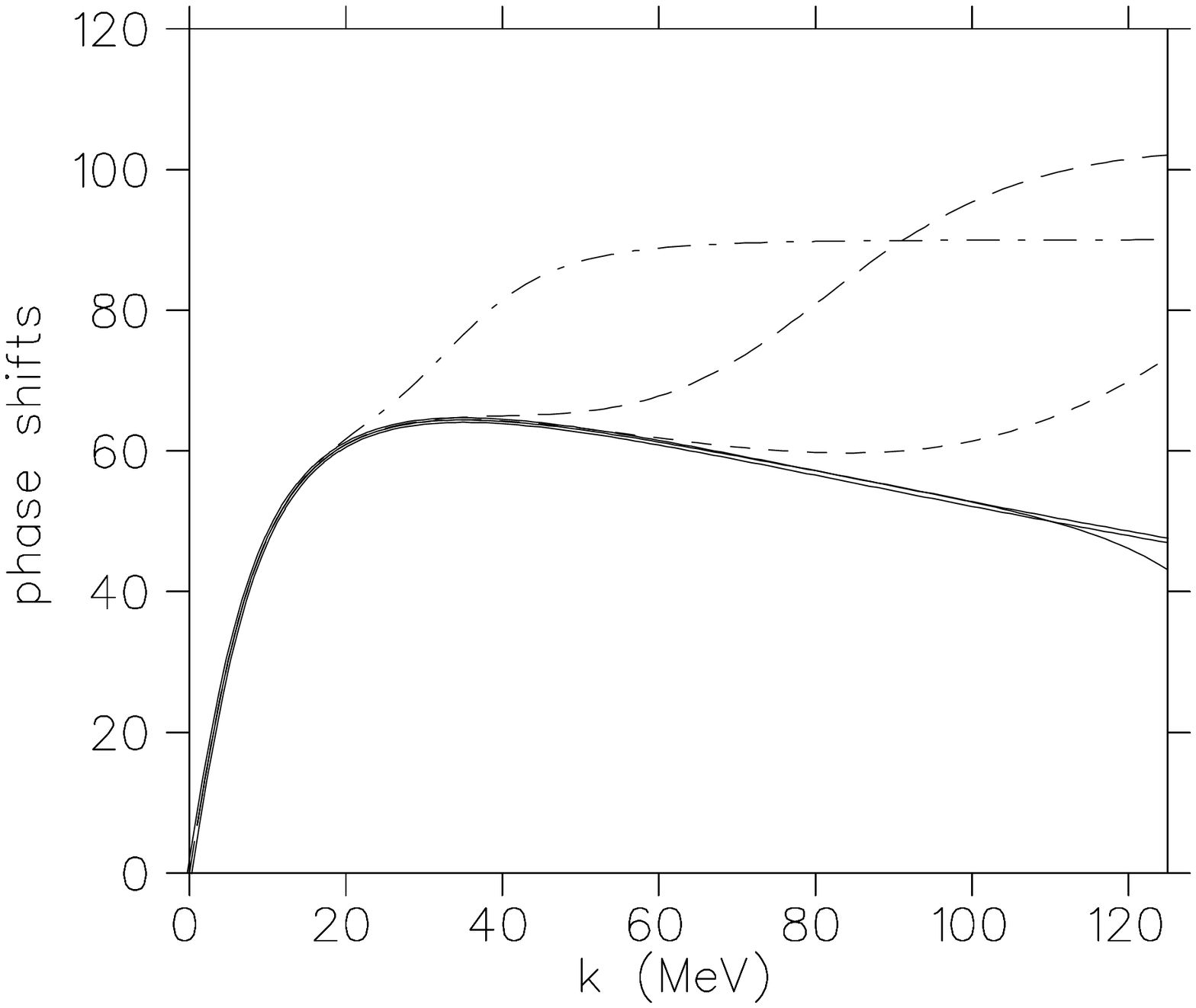}
\vspace{2mm}
\caption{\fign{phi4d}{\it Phase shifts calculated in $p^4$ order. Double line
corresponds to the effective
range expansion, solid line corresponds to the cut-off theory and 
dash-doted, long-dashed and short-dashed lines correspond to $\mu =0$, 
$\mu =40$ and
$\mu =130$ MeV respectively}}
\end{figure}

Substituting the numerical values for $a=-1/(8.4 \ {\rm MeV})$, $r_e=0.0137
\ {\rm MeV}^{-1}$, $d=0$ (the
first two terms in effective range expansion describe experimental data quite
well so there is no need to determine $d$ from data at least for the purposes of
this paper) 
and $l=130 \ {\rm MeV}$ one can calculate
coupling constants 
$\bar C\approx -{1/ (76.8 \ {\rm MeV})^2}$, 
$\bar C_2\approx {1/ (135.2 \ {\rm MeV})^4}$ (the sign ``-''
in (\ref{matchingcutpar2}) 
is again chosen respecting the structure of ordinary perturbation theory ),
$\bar B_1\approx -{1/ (124.6 \ {\rm MeV})^6}$. Using these
values one calculates phase
shifts from (\ref{b0t3}). These phase shifts are compared with results of
effective range
expansion and also of (\ref{rentmp4}) in
FIG. 3. The phase shifts of cut-off theory agree well with ones obtained from
subtractively renormalised theory for all momenta for which the second approach
describes the data well. One did not expect the results of
two approaches agree well beyond this range.

In FIG.4 the phase shifts for different  values of the cut-off
parameter  are plotted. It can be seen that for cut-offs $\sim m_\pi$ phase
shifts are cut-off independent up to $\sim 60 \ {\rm MeV}$.

\begin{figure}[t]
\hspace*{5cm}  \epsfxsize=8cm\epsfbox{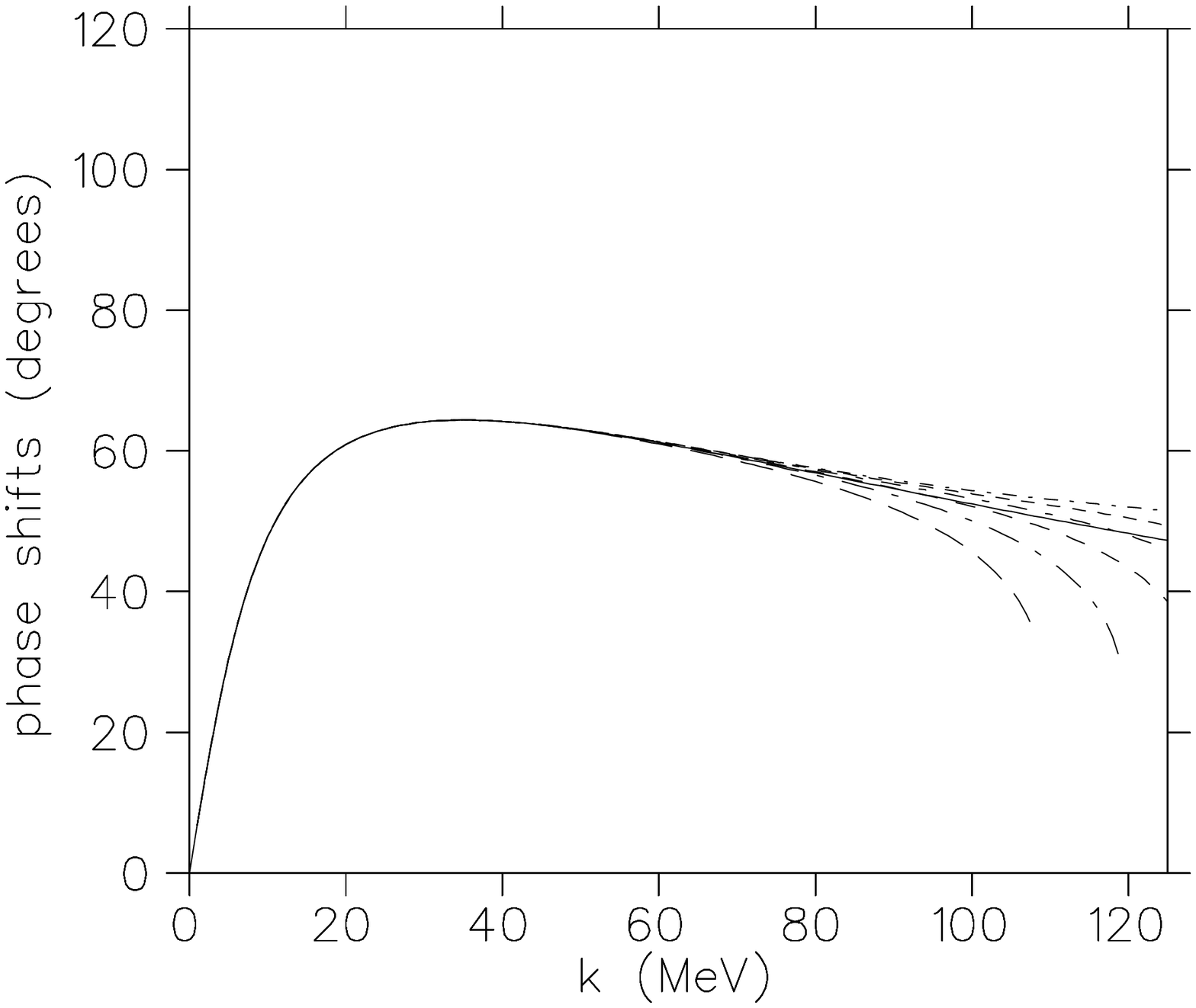}
\vspace{2mm}
\caption{\fign{cutshedareba} {\it Phase shifts calculated in $p^4$ order cut-off
theory.  Solid line corresponds to the effective range expansion. The lowest
dashed line corresponds to cut-off parameter $l=110 \ {\rm MeV}$ and subsequent
lines correspond to the values
$l=120, \ 130, \ 140, \ 150, \ 160, \ 170 \ {\rm MeV}$.}} 
\end{figure}

\section{Conclusions}
 
In the simple example of low energy effective field
theory for nucleon-nucleon
scattering it was demonstrated that the cut-off theory can reproduce the
scattering amplitude 
calculated using subtractively renormalised (using off mass-shell
subtractions) theory up to the order of accuracy of the
considered approximation, the difference between two amplitudes being of higher
order (in small expansion parameter). This simple
example serves as  a demonstration of some more general considerations about
cut-off field theories formulated below.

Using chiral power counting originally developed by Weinberg one can find the
potential up to any desired
order. Then to remove divergences one can impose cut-off regularization. 
The cut-off regularization destroys chiral and gauge symmetries and to restore
them it is necessary to include additional terms into the Lagrangian (and
consequently into the potential). Cut-off
dependence of the physical quantities can be removed systematically by including
additional terms in the Lagrangian \cite{lepage}. 
 
 The power-law divergences, which caused higher order
operators to be involved in the renormalization of the diagrams 
obtained by iterating the low order potential, now emerge as powers of the
cut-off parameter. As far as
cut-off should be taken of the order of masses of particles  which were
integrated out, it should be clear that cut-off regularization does not respect
power counting and it seems that imposing this regularization will destroy
the whole machinery. (The problem cannot be solved by imposing a small cut-off
as cut-off regularized integrals contain inverse powers of cut-off parameters as
well). 
However the large factors which seem to threaten power
counting can be absorbed by redefining the couplings already included  into the
potential \cite{park,lepage1}.

Fitting the parameters of
the cut-off theory one can reproduce
the results of the subtractively renormalised theory up to the order
of accuracy determined by approximation made in the potential. The results of
cut-off theory are as reliable as the ones of subtractively 
renormalised theory, the error being of the order of terms
neglected in the potential. As far as the cut-off is of the order of the mass of
lightest particle which was integrated out the higher order (in momenta)
cut-off dependent
corrections are suppressed by powers of this parameter. By increasing the
cut-off parameter one could make the
mentioned corrections smaller but for large cut-off it would be problematic to
include the positive powers of the cut-off parameter in to a redefinition of the
coupling constants. So, the equivalence between subtractively renormalized
and cut-off theories can be 
achieved only for the cut-off of the order of  the mass of lightest particle
which was integrated out.  
The results of two approaches are expected to deviate significantly beyond the
range of validity of the power counting. Although the difference
between two expressions is of higher order it becomes large
for the momenta for which the power counting breaks down in subtractively
renormalised theory. 
The range of validity of power counting depends on normalisation condition. 
Hence the range of the equivalence i.e. the range of the momenta for which the
difference between results of two approaches is small is determined by the
choice of normalisation condition in subtractively renormalised theory as well
as by the choice of the value of cut-off parameter in cut-off
theory. Choosing optimal conditions one can claim that two approaches lead to
equivalent results for the range of the momenta (energy) for which the power
counting is at work.

One should conclude that the doubts about consistency and systematic character
of
cut-off theories \cite{phillipswsh,epelbaoum,epelbaoum1} are ungrounded.
So the reasonable success of the cut-off chiral
perturbation theory originally started with work \cite{ordonez} should not be a
surprise. Although the cut-off theory is technically a little
complicated it has a
great advantage in that one can determine amplitudes from equations. Note that
there  are no
self-contained equations for subtractively renormalised amplitudes in these
non-renormalizable (in the traditional sense) effective field theories and
one instead has to sum up renormalised diagrams.

{\bf ACKNOWLEDGEMENTS}

I would like to thank B.Blankleider and A.Kvinikhidze for useful discussions and
N.Clisby for commenting on the manuscript.

This work was carried out whilst the author was a recipient of an 
 Overseas Postgraduate
Research Scholarship and a Flinders University Research Scholarship
at the Flinders University of South Australia.


\end{document}